\documentclass{aa}
 \usepackage{graphicx}
\usepackage[varg]{txfonts}
\graphicspath{{./Figures/}}

\begin{document}

\title{Formation of rocky and icy planetesimals inside and outside the snow line: Effects of diffusion, sublimation and back-reaction}

\author{Ryuki Hyodo\inst{\ref{inst1}} \and Shigeru Ida\inst{\ref{inst1}} \and S{\'e}bastien Charnoz\inst{\ref{inst2}}}
\institute{Earth-Life Science Institute/Tokyo Institute of Technology, 2-12-1 Tokyo, Japan \email{hyodo@elsi.jp} \label{inst1} \and Institut de Physique du Globe/Universit{\'e} Paris Diderot 75005 Paris, France\label{inst2}}

\abstract 
{Streaming instability is a possible mechanism to form icy planetesimals. It requires special local conditions such as a high solid-to-gas ratio at the midplane and typically more than centimeter size (Stokes number $> 0.01$). Silicate grains cannot grow to such a size through pairwise collisions. It is important to clarify where and when rocky and icy planetesimals are formed in a viscously evolving disk.}
{We wish to understand how local runaway pile-up of solids (silicate, water ice) occurs inside or outside the snow line.}
{We assume an icy pebble contains micron-sized silicate grains that are uniformly mixed with ice and are released during the ice sublimation. Using a local one-dimensional code, we solve the radial drift and the turbulent diffusion of solids and the water vapor, taking account of their sublimation/condensation around the snow line. We systematically investigate effects of back-reactions of the solids to gas on the radial drift and diffusion of solids, scale height evolution of the released silicate particles, and possible difference in effective viscous parameters between that for turbulent diffusion ($\alpha_{\rm tur}$) and that for the gas accretion rate onto the central star ($\alpha_{\rm acc}$). We also study the dependence on the ratio of the solid mass flux to the gas ($F_{\rm p/g}$).}
{We show that the favorable locations for the pile-up of silicate grains and icy pebbles are the regions in the proximity of the water snow line inside and outside it, respectively.  We found that runaway pile-ups occur when both the back-reactions for radial drift and diffusion are included. In the case with only the back-reaction for the radial drift, no runaway pile-up is found except for extremely high pebble flux, while the condition of streaming instability can be satisfied for relatively large $F_{\rm p/g}$ as found in the past literatures. If the back-reaction for radial diffusion is considered, the runaway pile-up occurs for a reasonable value of pebble flux. The runaway pile-up of silicate grains that would lead to formation of rocky planetesimals occurs for $\alpha_{\rm tur} \ll \alpha_{\rm acc}$, while the runaway pile-up of icy pebbles is favored for $\alpha_{\rm tur} \sim \alpha_{\rm acc}$. Based on these results, we discuss timings and locations of rocky and icy planetesimals in an evolving disk.}
{} 
\keywords{planets and satellites: formation $-$ planet-disk interactions $-$ accretion, accretion disks $-$ methods: numerical}

\titlerunning{Formation of rocky and icy planetesimals inside and outside the snow line}
\authorrunning{R. Hyodo, S. Ida and S. Charnoz}
 \maketitle
 
\section{Introduction} \label{sec:intro}
Planetesimals $-$ sub-km to several-hundred-km $-$ are thought to be the fundamental building blocks of all planets and small bodies in planetary systems. The "late-stage" planet formation would occur through the successive accumulation of planetesimals into planets \citep[e.g.][]{Saf72,Hay85}. On the other hand, the "early-stage" of planet formation would take place from micron-sized dust to planetesimals within the gas and dust disks called protoplanetary disks.\\

However, we are facing with a problem to link these "early-" and "late-"stages of planet formation because of the limitation of our understanding on planetesimal formation due mainly to two theoretical challenges: (1) so-called "growth barrier" which $\sim$cm-sized particles are too large to grow further due to fragmentation or bouncing during their high-speed collisions \citep{Blu00,Zso10} and (2) so-called "radial drift barrier" which the radial drift due to the gas drag of meter-sized particles is too fast for the particles to grow to planetesimals before they fall onto the host star \citep{Whi72,Wei77}. Thus, it seems difficult to form km-sized planetesimals from micron-sized dust through all intermediate sizes in a step-by-step manner. Note that \cite{Oku12} proposed that fluffy icy dust aggregates can grow to icy planetesimals without suffering from those barriers, but it is not certain if dust aggregates have such a fluffy structure.\\

A possible mechanism to form planetesimals directly from small particles is the so-called "streaming instability" (hereafter SI). SI is caused by momentum feedback of pebbles to the gas that leads to pebble clumping, and subsequent gravitational collapse is much faster than radial drift \citep{You05,Joh07a,Joh07b}. The streaming instability requires special conditions $-$ a high solid-to-gas ratio for sufficiently large pebbles \citep{Joh09,Bai10,Car15,Yan17}. In the previous works, many authors pointed out that the region outside the water snow line is a favorable location for SI, because water vapor released inside the snow line diffuses outward to re-condense outside the snow line and the recycling enhances the solid-to-gas ratio there \citep{Ste88,Ros13,Arm16,Sch17,Dra17,Sch18}. \\

While icy planetesimal formation just outside the snow line was discussed by many authors, rocky (silicate) planetesimals formation due to radial accumulation has not been discussed except in a few papers. The water faction of the terrestrial planets is small $-$ the ocean mass of the Earth is only $\sim0.023$ wt.\% of the total mass and a water mass preserved in Earth's mantle in a form of hydrous minerals is limited by $\sim$ the ocean mass \citep{Ber03,Hir06,Fei17}, while Earth's core could have $H$ equivalent to 2 wt.\% of $H_2O$ of the Earth \citep{Nom14}. Ancient Mars may have had $10^{-2} - 10^{-1}$ wt.\% water \citep{diA10,Cli10,Kur14}. Ancient Venus may also have had $10^{-3} - 10^{-1}$ wt.\% water and would have later lost through runway greenhouse effect \citep{Don82,Gre18}. These observations suggest that the building blocks of the terrestrial planets $-$ at least planetary embryos that underwent subsequent growth due to pebble accretion $-$ must be dry planetesimals with negligible water fraction \citep{Ida19}.\\

One possible idea for rocky planetesimal formation is the pile-up of silicate particles at a pressure bump such as the inner edge of the dead zone of magneto-rotational fluid instability \citep[e.g.][]{Ued19}. Another idea is the pile-up just inside the snow line. If $\mu$m to mm sized silicate grains are uniformly mixed within an icy pebble, they are released from progressively sublimating pebbles after passing the snow line. Because they are strongly coupled to the disk gas, they would be piled up just inside the snow line \citep{Sai11}. \\

While \cite{Sai11} assumed static gas disks, \cite{Ida16} considered accretion disks and found through a simple 1D analytical model that a runaway pile-up that would trigger "gravitational instability" (hereafter GI) is caused by the effect of slowing down of radial velocity of the silicate grains due to their pile-up ("back-reaction"). \cite{Ida16} reported that GI potentially occurs if pebble to gas mass flux ($F_{\rm p/g}$) is $\ga 0.3$ for $\alpha_{\rm tur}=10^{-3}$ (and lower value of $\alpha_{\rm tur}$ requires a lower value of $F_{\rm p/g}$ for GI to operate). However, their analytical argument neglected the radial diffusion of solids and vapor. Using a  more detailed 1D numerical simulation, \cite{Sch17} investigated the pile-up of icy pebbles outside the snow line and that of silicate grains inside the snow line, including the radial diffusion associated with the turbulent diffusion of disk gas $-$ which was not included in \cite{Ida16}. They found that the water vapor recycling due to radial diffusion causes a high enough solid-to-gas ratio for SI beyond the snow line, but find neither significant pile-up nor the runaway pile-up of silicate grains inside the snow line.\\

\cite{Sch17} included the back-reaction that slows the radial drift of icy pebbles as they pile-up, which facilitates the occurrence of SI.  But, they assumed that the radial drift of silicate grains is unaffected by their pile-up. It suggests that the back-reaction for the silicate grains may play an essential role in the runaway pile-up found by \cite{Ida16}.  We also note that \cite{Ida16} neglected that the vertical stirring of the silicate grains and assumed their scale height is the same as that of the incident icy pebbles, which also enhances the runaway pile-up, while \cite{Sch17} assumed that the silicate grains are instantaneously mixed with the background gas $-$ that is, silicate grains have the same scale height at the gas. In reality, the silicate grain scale height is the same as the icy pebbles when the grains are released from the sublimating icy pebbles and it is gradually increased by the vertical turbulent stirring as the grains migrate inward from the snow line. Thus, these two studies have different assumptions and models. So, it is still not fully understood if the runaway pile-up of silicate grains inside the snow line occurs or not if all of the effects of the back-reaction, the radial diffusion, and the evolution of the silicate grains' scale height are incorporated. \\

Here, following the approach by \cite{Sch17}, we further investigate the disk evolution and the pile-up of solids (silicate grains and icy pebbles) inside and outside of the water snow line by considering several effects that were neglected in the previous work. We additionally consider (1) the back-reaction of silicate grains onto the gas that slows radial drift of the silicate grains, while only the back-reaction of pebbles are considered by \cite{Sch17}, (2) two different back-reactions of solids onto the gas $-$ we call them the radial drift back-reaction that reduces radial drift velocity of solids (hereafter "Drift-BKR") and the diffusion back-reaction that weakens turbulent diffusion of solids (hereafter "Diff-BKR"), which has not taken into account before, respectively and (3) the scale height of the released silicate grains that is initially equal to that of icy pebbles at the snow line and is increased by turbulent stirring up to the gas scale height. We also distinguish (4) two different effective viscous parameters $-$ one that regulates accretion of the gas to the host star, $\alpha_{\rm acc}$, and the other that regulates the midplane diffusion of solids, $\alpha_{\rm tur}$. The previous work assumed $\alpha_{\rm acc}=\alpha_{\rm tur}$.  We will show that Drift-BKR and Diff-BKR of silicate grains are the most essential factor for the occurrence of the runaway pile-up of silicate grains and that of icy pebbles is also resulted in by Diff-BKR.  The ratio between $\alpha_{\rm tur}$ and $\alpha_{\rm acc}$ controls relative importance between the runaway pile-up of silicate grains inside the snow line and that of icy pebbles outside the snow line. We map the parameter regions of $F_{\rm p/g}$ and $\alpha_{\rm tur}/\alpha_{\rm acc}$ for GI (runaway pile-up) of silicate grains, GI of icy pebbles and SI of the icy pebbles.\\

In section \ref{sec_methods}, we describe our basic set-ups of numerical approaches. In section \ref{sec_results}, we present our numerical results. In section \ref{sec_discussion}, we discuss the results. In section \ref{sec_summary}, we summarize our paper.\\

\section{Numerical methods}
\label{sec_methods}

\subsection{Gas Disk Model}
In this paper, we use classical $\alpha$-accretion disk model \citep{Sha73,Lyn74} where the surface density of the gas is expressed as a function of the gas accretion rate $\dot{M}_{\rm gas}$ and the effective viscosity $\nu_{\rm acc}$ as
\begin{equation}
\label{eq_sigma}
	\Sigma_{\rm gas}=\frac{\dot{M}_{\rm gas}}{3\pi \nu_{\rm acc}}
\end{equation}
and $\nu_{\rm acc}$ is written using the sound speed $c_{\rm s}$, the Keplerian orbital frequency $\Omega$ and the dimensionless effective viscous parameter $\alpha_{\rm acc}$ given as
\begin{equation}
	\nu_{\rm acc}=\alpha_{\rm acc}c_{\rm s}^2 \Omega^{-1}.
\end{equation}
The above equation assumes a steady state of the background gas. When the gas accretion is dominated by the viscous diffusion, $\alpha_{\rm acc}$ equals the dimensionless turbulence parameter $\alpha_{\rm tur}$. In contrast, when the gas accretion is regulated by the disk wind-driven accretion \citep{Bai16,Suz16}, generally $\alpha_{\rm acc}$ is larger than $\alpha_{\rm tur}$ \citep{Arm13,Has17}.\\

The isothermal sound speed of the gas is written as
\begin{equation}
	c_{\rm s} = \sqrt{ \frac{k_{\rm B} T}{\mu_{\rm gas}}}
\end{equation}
where $k_{\rm B}$ is the Boltzmann constant, $\mu_{\rm gas}$ is the molecular weight of the gas and $T$ is the temperature of the gas. In this paper, the temperature profile is fixed as
\begin{equation}
	T=150 \left( \frac{r}{3.0 {\rm au} } \right)^{-1/2} {\rm K}.
\end{equation}
Note that our disk is somewhat a hot disk but we chose this profile so that we can directly compare with the previous work \citep{Sch17}.\\

Under the above conditions, the gas scale height is written as
\begin{equation}
\label{eq_H_gas}
	H_{\rm gas}=\frac{c_{\rm s}}{\Omega} \sim 0.033 \left( \frac{r}{1.0 {\rm au} } \right)^{5/4} {\rm au}.
\end{equation} 
The midplane gas density is $\rho_{\rm gas}=\Sigma_{\rm gas}/\sqrt{2\pi}H_{\rm gas}$. In this paper, we assume that the above gas profile is unchanged through the simulation and thus $\mu_{\rm gas}=2.34m_{\rm proton}$, where $m_{\rm proton}$ is the proton mass. The accretion velocity of the gas $v_{\rm gas}$ is written as 
\begin{equation}
	v_{\rm gas} = -\frac{3}{2r} \nu_{\rm acc}.
\end{equation}
where a negative value indicates radial drift towards the central star.\\

\subsection{Radial motions of solids and vapor}
In this work, we describe a pebble as the mixture of many micron-sized silicate grains covered by water ice that is also used in the previous paper (see "many-seeds pebble model" in \cite{Sch17} and the models used by \cite{Sai11} and \cite{Ida16}). In our one-dimensional calculation,
we adopt the single-size approximation \citep{Orm14,Sat16} where the pebble size is represented by a single size depending on the radial distance $r$, and solve the system integrated over the vertical direction of the disk assuming pebbles and silicate grains have a Gaussian distribution in the vertical direction whose scale heights are $H_{\rm peb}$ and $H_{\rm sil}$, respectively.\\

The Stokes number of a particle ${\rm St}=t_{\rm s}\Omega$ is written by using the stopping time which represents the relaxation timescale of particle momentum through the gas drag as
\begin{equation}
	t_{\rm s} = \frac{\rho_{\rm par}r_{\rm peb}}{v_{\rm th}\rho_{\rm gas}} \hspace{2em} {\rm Epstein:} s_{\rm peb} < \frac{9}{4} \lambda_{\rm mfp} 
\end{equation}
\begin{equation}
	t_{\rm s} = \frac{4\rho_{\rm par}r_{\rm peb}^2}{9\rho_{\rm gas}v_{\rm th}\lambda_{\rm mfp}} \hspace{2em} {\rm Stokes:} s_{\rm peb} > \frac{9}{4} \lambda_{\rm mfp} 
\end{equation}
where $v_{\rm th}=\sqrt{8/\pi}c_{\rm s}$ is the thermal velocity, $\lambda_{\rm mfp}=\mu/\sqrt{2}\rho_{\rm gas}\sigma_{\rm mol}$ is the mean free path of the gas, $r_{\rm peb}$ is the size of a pebble, $\sigma_{\rm mol}=2.0\times10^{-15}$ cm$^{2}$ and $\rho_{\rm par}$ and $\rho_{\rm gas}$ are the particle internal density and the gas spatial density, respectively.\\

Pebbles and silicate grains orbit within the gas whose rotation is sub-Keplerian due to the pressure gradient. Thus, angular momentums of pebbles and silicate grains are lost through the gas drag and they spiral inward towards the star. When an icy pebble approaches its snow line, water ice starts sublimating and water vapor is produced. As we assume that micron-sized silicate grains are uniformly mixed within a pebble, the silicate grains are also released as the water vapor sublimates (see more details in the next section). The governing equations of the surface density of icy component within a pebble $\Sigma_{\rm ice,peb}$, that of silicate component within a pebble $\Sigma_{\rm sil,peb}$, that of released silicate grains $\Sigma_{\rm sil,gas}$ and that of water vapor $\Sigma_{\rm vap}$ as well as the number density of pebbles $N_{\rm peb}$ are given as \citep{Des17} 
\begin{equation}
\label{eq_sigma_ice_peb}
	\frac{\partial \Sigma_{\rm ice,peb}}{\partial t} + \frac{1}{r}\frac{\partial}{\partial r} \left( r \Sigma_{\rm ice,peb} v_{\rm peb} - r D_{\rm peb}\Sigma_{\rm gas} \frac{\partial}{\partial r} \left( \frac{\Sigma_{\rm ice,peb}}{\Sigma_{\rm gas} } \right) \right) =\dot{\Sigma}_{\rm ice,peb}
\end{equation}
\begin{equation}
	\frac{\partial \Sigma_{\rm sil,peb}}{\partial t} +  \frac{1}{r}\frac{\partial}{\partial r}  \left( r \Sigma_{\rm sil,peb} v_{\rm peb} - r D_{\rm peb}\Sigma_{\rm gas}  \frac{\partial}{\partial r}  \left( \frac{\Sigma_{\rm sil,peb}}{\Sigma_{\rm gas} } \right) \right) =\dot{\Sigma}_{\rm sil,peb}
\end{equation}
\begin{equation}
	\frac{\partial \Sigma_{\rm sil,gas}}{\partial t} +  \frac{1}{r}\frac{\partial}{\partial r}  \left( r \Sigma_{\rm sil,gas} v_{\rm sil} - r D_{\rm sil}\Sigma_{\rm gas}  \frac{\partial}{\partial r}  \left( \frac{\Sigma_{\rm sil,gas}}{\Sigma_{\rm gas} } \right) \right) =\dot{\Sigma}_{\rm sil,gas}
\end{equation}
\begin{equation}
\label{eq_sigma_vap}
	\frac{\partial \Sigma_{\rm vap}}{\partial t} +  \frac{1}{r}\frac{\partial}{\partial r}  \left( r \Sigma_{\rm vap} v_{\rm gas} - r D_{\rm gas}\Sigma_{\rm gas}  \frac{\partial}{\partial r}  \left( \frac{\Sigma_{\rm vap}}{\Sigma_{\rm gas} } \right) \right) =\dot{\Sigma}_{\rm vap}
\end{equation}
\begin{equation}
\label{eq_Np}
	\frac{\partial N_{\rm peb}}{\partial t} +  \frac{1}{r}\frac{\partial}{\partial r}  \left( r N_{\rm peb} v_{\rm peb} - r D_{\rm peb}N_{\rm gas}  \frac{\partial}{\partial r}  \left( \frac{N_{\rm peb}}{N_{\rm gas} } \right) \right) = 0
\end{equation}
where $v_{\rm peb}$, $v_{\rm sil}$ and $v_{\rm gas}$ are the radial velocity of pebbles, that of silicate grain and that of the gas, respectively. $N_{\rm gas}$ is the number density of the gas. Source terms on the right-hand sides of Eqs. (\ref{eq_sigma_ice_peb})-(\ref{eq_sigma_vap}) are due to sublimation and condensation of ice in a pebble, and due to reaccretion of silicate grains onto icy pebbles as discussed in the following section and the sum of these equations become zero. Sublimation/condensation occurs only when a pebble exists and thus the right-hand side of Eq. (\ref{eq_Np}) is zero. $D_{\rm peb}$, $D_{\rm sil}$ and $D_{\rm gas}$ are the diffusivities of pebbles, silicate grains and vapor and we will discuss in more details in the next section.\\

The midplane densities of pebbles and silicate grains are written as $\rho_{\rm peb}=\Sigma_{\rm peb}/\sqrt{2\pi}H_{\rm peb}$ and $\rho_{\rm sil}=\Sigma_{\rm sil}/\sqrt{2\pi}H_{\rm sil}$, where $H_{\rm peb}$ and $H_{\rm sil}$ are the scale heights of pebbles and silicate grains, respectively. The scale height of pebbles is assumed to be
\begin{equation}
\label{eq_H_pebble}
	H_{\rm peb}=H_{\rm gas}\sqrt{ \frac{\alpha_{\rm tur}} {{\rm St}_{\rm peb} + \alpha_{\rm tur}}}
\end{equation}
where $\alpha_{\rm tur}$ is the dimensionless turbulent parameter for diffusion. In contrast, we assume that the released silicate grains from icy pebbles initially have the same scale height as that of pebbles. It increases up to that of the gas by vertical turbulent diffusion. We model the time evolution of the scale height of silicate grains as
\begin{equation}
\label{eq_silicate_H}
	H_{\rm sil}=H_{\rm gas}  \left( 1+ \frac{ {\rm St}_{\rm peb} }{\alpha_{\rm tur}} \times e^{ \frac{-\Delta t}{t_{\rm mix}}} \right)^{-1/2}
\end{equation}
where $\Delta t=|(r-r_{\rm snow})/v_{\rm sil}|$ and $r_{\rm snow}$ is the radial distance of the snow line to the star. $t_{\rm mix}=(H_{\rm gas}/l_{\rm mfp})^2/\Omega$ is the diffusion/mixing timescale to the vertical direction at the snow line and $ l_{\rm mfp}=\sqrt{\alpha_{\rm tur}}H_{\rm gas}$ is the  the mean free path of turbulent blobs. Here $\Delta t$ and $t_{\rm mix}$ are calculated by using the physical values at the snow line and ${\rm St_{\rm peb}}$ is the Stokes number of pebbles at the snow line.\\

Through the arguments of a static disk limitation \citep{Nak86} and a low solid-to-gas ratio limitation \citep{Gui14}, the radial drift velocity of the gas is affected by the back-reaction of the solids to the gas. As the gas velocity changes through the back-reaction of solids, the motions of the solids are correspondingly affected as a result of exchanging the momentum with the gas. Thus, the radial drift velocity of a pebble and a silicate grain are given as \citep[see also][]{Ida16,Sch17}
\begin{equation}
\label{eq_BR_solid1}
	v_{\rm peb} = - \frac{2\eta v_{\rm K}{\rm St}_{\rm peb} }{ {\rm St}^2_{\rm peb} + (1+Z_{\rm peb})^2} + \frac{1+Z_{\rm peb}}{ {\rm St}^2_{\rm peb} + (1+Z_{\rm peb})^2} v_{\rm gas}
\end{equation}
\begin{equation}
\label{eq_BR_solid2}
	v_{\rm sil} = - \frac{2\eta v_{\rm K}{\rm St}_{\rm sil}}{ {\rm St}^2_{\rm sil} + (1+Z_{\rm sil})^2} + \frac{1+Z_{\rm sil}}{ {\rm St}^2_{\rm sil} + (1+Z_{\rm sil})^2} v_{\rm gas}
\end{equation}
where $Z_{\rm peb}=\rho_{\rm peb}/\rho_{\rm gas}$ and $Z_{\rm sil}=\rho_{\rm sil}/\rho_{\rm gas}$ are the midplane solid-to-gas density ratios of pebbles and silicate grains, respectively. ${\rm St}_{\rm peb}$ and ${\rm St}_{\rm sil}$ are the Stokes numbers of pebbles and silicate grains, respectively. $\eta v_{\rm K}$ is the maximum radial drift speed owing to the gas pressure gradient to the radial direction, given by
\begin{equation}
	\eta v_{\rm K} = -\frac{1}{2} (c_{\rm s}^2/v_{\rm K}) \frac{\partial \log P}{\partial \log r}
\end{equation}
where $v_{\rm K}$ is the Keplerian velocity and $P$ is the midplane volume-valued total pressure including the effect of both the gas and water vapor produced around the snow line.\\\\

We assume that the background gas is unaffected as in the previous papers \citep{Ida16,Sch17} and we only consider the effects of the back-reaction onto the motions of solids as described in Eqs. (\ref{eq_BR_solid1}) and (\ref{eq_BR_solid2}). Note that the same effect of the back-reaction that changes the velocity of solids is referred to in different ways in different papers. For example, \cite{Ida16} refers to it as "the back-reaction of the gas on the motion of solid" while \cite{Sch17} refers to it as "the back-reaction of solids on the gas". In both cases, they consider the change of radial drift of solids as they pile up. In this work, we refer to the back-reaction that changes the motion of the solids as "back-reaction of solids to the gas". \\

\subsection{Diffusions of pebbles and silicate grains}
In this study, we study the importance of diffusion on the local pile-up of solids. First, we define the gas/vapor diffusivity $D_{\rm gas}$, given by
\begin{equation}
\label{eq_D_gas}
	D_{\rm gas} = \nu_{\rm tur} = \alpha_{\rm tur}c_{\rm s} H_{\rm gas}
\end{equation}
where $\nu_{\rm tur}$ is the turbulent viscosity that regulates diffusion in association with dimensionless turbulence parameter, $\alpha_{\rm tur}$.\\

The nature of solid diffusion in the gas is partial coupling with the gas eddies and its diffusivity $D_{\rm solid}$ is often related to $D_{\rm gas}$ through the Schmidt number $\mathrm{Sc}$ \citep{You07} as
\begin{equation}
\label{eq_D_solid}
	D_{\rm solid} = D_{\rm gas} \times \mathrm{Sc} = \frac{D_{\rm gas}}{1+{\rm St}^2}.
\end{equation}
\\

When local pile-up of pebbles or silicate grains occurs, the diffusivity should also be affected by its collective effect on the turbulent eddies. Diffusivity has a dimension of the square of velocity multiplied by a characteristic timescale of the system. We note that the dependence of the radial drift velocity under Drift-BKR (and thus the dependence of the velocity) on the solid's pile-up  is $1/(1+Z)^2$ (see Eqs. (\ref{eq_BR_solid1}) and (\ref{eq_BR_solid2})). Thus, we assume that the diffusivity has the same dependence on $1/(1+Z)^2$. However, this is just a simple consideration and the actual dependence on $Z$ is not clear. Thus, in order to assess its dependence on the results, we also study the case when the diffusivity depends on $1/(1+Z)$. Thus, we describe the diffusivity as follow
\begin{equation}
\label{eq_D_pebble}
	D_{\rm peb} = \frac{D_{\rm gas}}{1+{\rm St_{\rm peb}}^2} \times \left( \frac{1}{1+Z_{\rm peb}} \right)^{K}
\end{equation}
\begin{equation}
\label{eq_D_silicate}
	D_{\rm sil} = \frac{D_{\rm gas}}{1+{\rm St_{\rm sil}}^2} \times \left( \frac{1}{1+Z_{\rm sil}} \right)^{K}
\end{equation}
where $K$ is the coefficient and we use $K=0$, $1$ or $2$. In this paper, we call these effects of the pile-ups of solids on their diffusivities as the radial diffusion back-reaction (Diff-BKR).\\

\subsection{Sublimation and condensation of water ice}
When an icy pebble approaches the water snow line, water ice starts to sublimate. As the micron-sized grain is assumed to be uniformly mixed within a pebble, the grains are assumed to be also released from a pebble at the same rate of vapor production (see also \cite{Sch17}).\\

In our calculation, we calculate the saturating vapor partial pressure described by the Clausius-Clapeyron equation:
\begin{equation}
	P_{\rm eq} = P_{\rm eq,0} e^{-T_{\rm 0}/T}
\end{equation}
where $P_{\rm eq,0}=1.14 \times 10^{13}$ g cm$^{-1}$ s$^{-2}$ and $T_0=6062$ K \citep{Lic91}. We also calculate water vapor pressure $P_{\rm vap}$, given by
\begin{equation}
	P_{\rm vap} = \frac{\Sigma_{\rm vap}}{\sqrt{2\pi} H_{\rm gas}} \frac{k_{\rm B}T}{\mu_{\rm H_2O}}
\end{equation}
where $\mu_{\rm H2O}=18m_{\rm proton}$ and we assume that water vapor instantaneously mixes with the background gas. We define the snow line is where $P_{\rm vap}$ equals $P_{\rm eq}$.\\

When sublimation/condensation occurs, the source terms of ice and water vapor are given by \citep{Sch17}
\begin{equation}
	\dot{\Sigma}_{\rm ice,peb} = (R_{\rm con} \Sigma_{\rm vap} - R_{\rm eva}) \Sigma_{\rm ice,peb}
\end{equation}
\begin{equation}
	\dot{\Sigma}_{\rm vap} = -(R_{\rm con} \Sigma_{\rm vap} - R_{\rm eva}) \Sigma_{\rm ice,peb}.
\end{equation}
where $R_{\rm eva}$ and $R_{\rm con}$ are
\begin{equation}
	R_{\rm eva} = 8 \sqrt{2\pi} \frac{r_{\rm peb}^2}{m_{\rm peb}} \sqrt{\frac{\mu_{\rm H_2O}}{k_{\rm B}T}} P_{\rm eq}
\end{equation} 
and 
\begin{equation}
	R_{\rm con} =  8 \sqrt{ \frac{k_{\rm B} T}{\mu_{\rm H_2O}}} \frac{r_{\rm peb}^2}{m_{\rm peb} H_{\rm gas}} 
\end{equation} 
respectively \citep{Sch17}.
\\

At the same time of ice sublimation, silicate grains are released at the same rate. Also, silicate grains can stick to an icy pebble at a rate of $R_{\rm s}$ \citep{Sch17}:
\begin{equation}
	R_{\rm s} = \frac{\Sigma_{\rm ice,peb} + \Sigma_{\rm sil,peb}}{\sqrt{2\pi}H_{\rm sil}m_{\rm peb}} \Delta v_{\rm sil,peb} \pi s_{\rm peb}^2
\end{equation}
where relative velocity between a pebble and a silicate grain is assumed to be $\Delta v_{\rm sil,peb}=v_{\rm peb}$. Thus, when pebbles sublimate, the source terms of pebbles and silicate grains are  
\begin{equation}
	\dot{\Sigma}_{\rm sil,peb} =  (R_{\rm con} \Sigma_{\rm vap} - R_{\rm eva}) \Sigma_{\rm sil,peb} + R_{\rm s}\Sigma_{\rm sil,gas} \\ 
\end{equation}
\begin{equation}
	\dot{\Sigma}_{\rm sil,gas} = -(R_{\rm con} \Sigma_{\rm vap} - R_{\rm eva}) \Sigma_{\rm sil,peb} - R_{\rm s}\Sigma_{\rm sil,gas} 
\end{equation}
and when pebbles condense,
\begin{equation}
\label{eq_silicate_stick}
	\dot{\Sigma}_{\rm sil,peb} =  R_{\rm s}\Sigma_{\rm sil,gas} \hspace{2em} 
\end{equation}
\begin{equation}
	\dot{\Sigma}_{\rm sil,gas} =  -R_{\rm s}\Sigma_{\rm sil,gas}.
\end{equation}
\\

\subsection{Numerical parameters and settings}
In this work, we use a local one-dimensional calculation between $r=0.1-5$ au divided by $1000-3000$ bins, depending on the parameters. In our model, a pebble at the outer boundary contains an equal mass of ice and silicate with its Stokes number of 0.1 at 3 au. The size of silicate grains are assumed to be micron-size ($\rm{St} \ll 0.01$). We set the mass of the central star to be one solar mass $M_{\rm sun}$. We study the dependence on the disk gas accretion rate ($\dot{M}_{\rm gas}=3\times10^{-9}$, $1\times10^{-8}$ and $3\times10^{-8} M_{\rm sun}$/year) and the dimensionless parameter for the radial accretion ($\alpha_{\rm acc}=3\times10^{-3}$ and $1\times10^{-2}$) and the dimensionless turbulence parameter for solid diffusion ($\alpha_{\rm tur}=1\times10^{-4}-1\times10^{-2}$ where $\alpha_{\rm tur} \leq \alpha_{\rm acc}$). We also change the ratio of pebble accretion rate $\dot{M}_{\rm peb}$ to gas accretion rate at the outer boundary as a parameter, given by
\begin{equation}
	F_{\rm p/g} = \frac{\dot{M}_{\rm peb}}{\dot{M}_{\rm gas}} = 0.1 - 0.6.
\end{equation}
The pebble surface density at the outer boundary is fixed as 
\begin{equation}
	\Sigma_{\rm peb,out} = F_{\rm p/g} \times \frac{\dot{M}_{\rm gas}}{2\pi r v_{\rm peb,out}}
\end{equation}
where $v_{\rm peb,out}$ is the velocity of pebbles at the outer boundary. Note that, as $\alpha_{\rm acc}$ increases, the gas surface density ($\propto 1/\alpha_{\rm acc}$) decreases but the pebble surface density ($\propto 1/v_{\rm peb,out}$) decreases only a small fraction $-$ This is because the drift of pebbles (${\rm St} \sim 0.1$) is mainly regulated by the gas drag (the first term of Eq. (\ref{eq_BR_solid1})) and not by the advection of the gas (the second term of Eq. (\ref{eq_BR_solid1})). Thus, as $\alpha_{\rm acc}$ increases, the midplane solid-to-gas ratio at the outer boundary increases. As $\alpha_{\rm tur}$ decreases, the scale height of pebbles decreases (Eq. (\ref{eq_H_pebble})) and thus the solid-to-gas ratio at the outer boundary increases.\\

\section{Numerical results}
\label{sec_results}
In this section, we show the results of our numerical simulations and discuss the effects of Drift-BKR (Eqs. (\ref{eq_BR_solid1}) and (\ref{eq_BR_solid2})) and Diff-BKR (Eqs. (\ref{eq_D_pebble}) and (\ref{eq_D_silicate})) on the pile-up of solids inside/outside the snow line. In section \ref{sec_K0}, we show the cases where Drift-BKR is included but Diff-BKR is not included. In section \ref{sec_SO17}, we show the cases where Drift-BKR of icy pebbles is included and Drift-BKR of silicate grains is neglected/included and we discuss the effects of Drift-BKR of silicate grains on the pile-up of solids. In section \ref{sec_steady_state} - \ref{sec_dep_accretion_rate}, we show the cases where Drift-BKR of both icy pebbles and silicate grains are included. In section \ref{sec_K1K2}, we discuss the cases where both Drift-BKR and Diff-BKR are included and we show that runaway pile-ups of solids inside/outside the snow line occurs for some reasonable range of disk parameters.\\

\subsection{Case without the diffusion back-reaction ($K=0$)}
\label{sec_K0}
\subsubsection{Importance of radial drift back-reaction (Drift-BKR) onto the motion of silicate grains}
\label{sec_SO17}
In \cite{Sch17}, Drift-BKR of pebbles was included but that of silicate grains was neglected. Here, we show that the Drift-BKR of silicate grains also plays an important role in the local pile-up of silicate grains. In section \ref{sec_K0}, Diff-BKR is neglected ($K=0$). \\
 
\begin{figure*}
\resizebox{\hsize}{!}{ \includegraphics{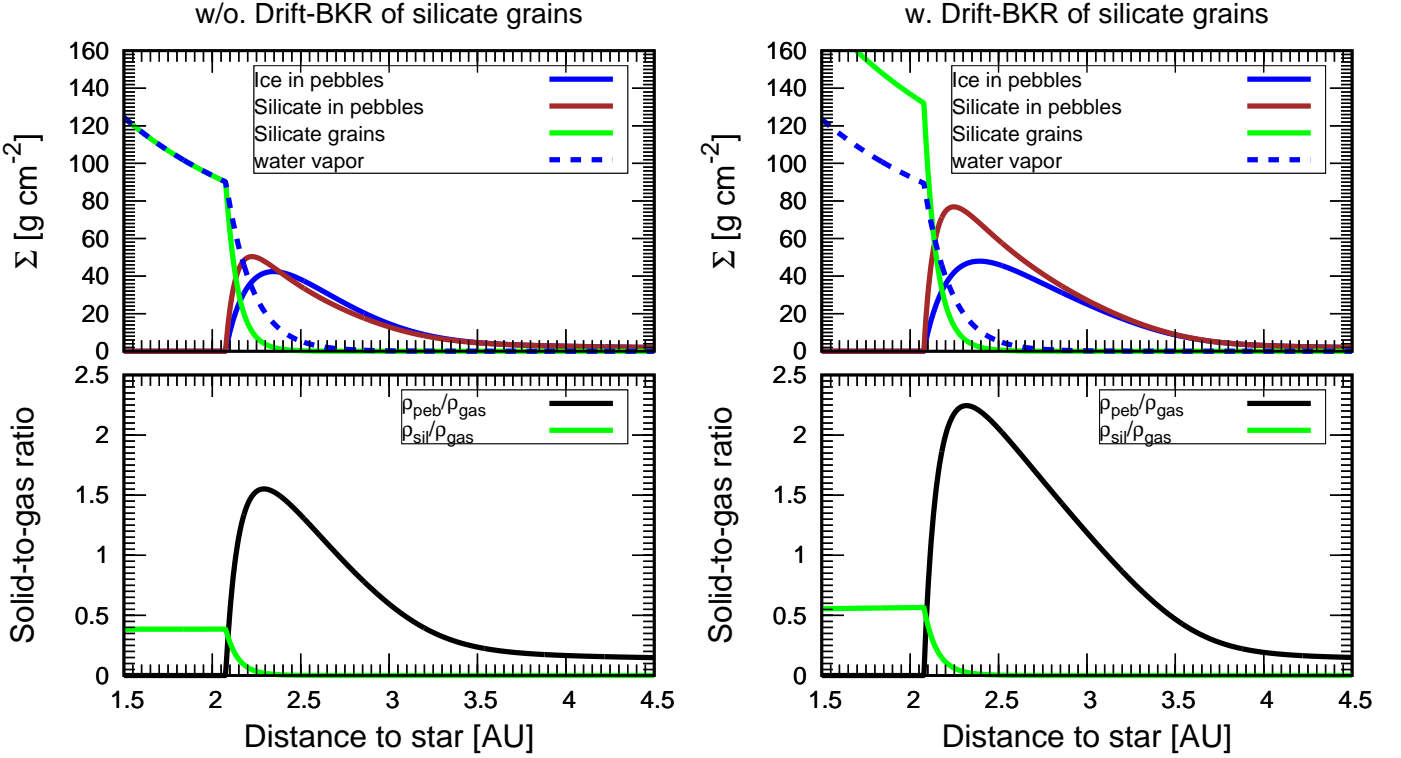}}
\caption{Surface density profile (top) and midplane solid-to-gas density ratio (bottom). Left panels show a case where Drift-BKR of only pebbles is considered (the same setting with \cite{Sch17}). Right panels show a case where Drift-BKR of both pebbles and the released silicate grains are considered. In the top panels, blue, brown and green solid lines represent those of ice in pebbles, silicate in pebbles and silicate grains, respectively. Dashed blue lines in the top panels represent those of water vapor. In the bottom panels, black and green lines represent those of pebbles and silicate grains, respectively. Here, we assume that the silicate grains are instantaneously mixed with the background gas and have the same scale height with the gas.  We also assume that $\alpha_{\rm acc}=\alpha_{\rm tur}=3 \times10^{-3}$, ${\rm St}=0.03$ at 3 au and $F_{\rm p/g}=0.8$ for the direct comparison between the left panels and the results of \cite{Sch17} ("many-seeds" model of Figure 5 in \cite{Sch17}).}
\label{SO17_comp}
\end{figure*}

In Figure \ref{SO17_comp}, we show the surface density profiles and the midplane dust-to-gas ratios for two different cases $-$ (1) the case where only Drift-BKR of pebbles are included (left panels in Figure \ref{SO17_comp}) and (2) the case where Drift-BKR of both pebbles and the released silicate grains are included (right panels in Figure \ref{SO17_comp}). Here, in order to directly compare with the previous study \citep{Sch17}, we assume that the scale height of the released silicate grains are instantaneously mixed with the background gas, $H_{\rm sil} = H_{\rm gas}$ (the same setting as \cite{Sch17}). We also assume that $\alpha_{\rm acc}=\alpha_{\rm tur}=3 \times10^{-3}$, ${\rm St}=0.03$ at 3 au and $F_{\rm p/g}=0.8$ for the direct comparison between the left panels and the results of \cite{Sch17} ("many-seeds" model of Figure 5 in \cite{Sch17}).\\

As pebbles approach near the snow line, they start to sublimate and water vapor and silicate grains are released from pebbles (dashed blue lines and green solid lines in the top panels of Figure \ref{SO17_comp}, respectively). A fraction of water vapor diffuses outward and re-condenses onto pebbles. Also, a fraction of silicate grains diffuses outward and sticks onto pebbles. Thus, just outside the snow line, pebbles pile up and their solid-to-gas ratio is enhanced. Our results (left panels of Figure \ref{SO17_comp}) well reproduce the results of \cite{Sch17} even though our pile-up of pebbles look slightly radially narrower than their results. When Drift-BKR of silicate grains is neglected (left panels of Figure \ref{SO17_comp}), water vapor and silicate grains (whose Stokes number is small enough for the grains to be well coupled to the gas) have almost the same drift velocity and the initial silicate-to-ice ratio of inflow from the outer boundary is unity. Thus, the surface densities of the water vapor and silicate grains are almost the same inside the water snow line. (left top panel of Figure \ref{SO17_comp}). Including Drift-BKR of silicate grains reduces their radial drift velocity and thus the pile-up of silicate grains inside the snow line is further enhanced (right panels of Figure \ref{SO17_comp}) compared to the case where Drift-BKR of silicate grains is neglected (left panels of Figure \ref{SO17_comp}). Outward diffusion of silicate grains is also enhanced and thus the fraction of the silicate component within pebbles outside the snow line is enhanced (right top panel of Figure \ref{SO17_comp}). Figure \ref{SO17_comp} clearly shows that including Drift-BKR of silicate grains increases not only the pile-up of silicate grains inside the snow line but also it enhances and widens the pile-up of pebbles outside of the snow line. Therefore, Drift-BKR of silicate grains has a non-negligible effect on the local pile-ups both inside and outside of the snow line.\\

\cite{Ida16} predicted a runaway pile-up of silicate grains for (their Eq. (18))
\begin{equation}
\label{eq_sil_runaway}
   F_{\rm p/g} > (Z_{\rm peb}/Z_{\rm sil})(H_{\rm sil}/H_{\rm gas}).
\end{equation}
They took account of Drift-BKR of silicate grains but neglected the radial diffusion. We also performed simulations with Drift-BKR of silicate grains of $\alpha_{\rm tur} = 10^{-4} - 10^{-3}$, $H_{\rm sil} = H_{\rm peb} = 0.03-0.1 H_{\rm gas}$ and $Z_{\rm peb}/Z_{\rm sil}=2$ at the outer boundary. We found that the threshold $F_{\rm p/g}$ is consistent with that predicted by Eq. (\ref{eq_sil_runaway}), although the simulation included the radial diffusion that tends to suppress the pile-up. If we adopt Eq. (\ref{eq_silicate_H}) for $H_{\rm sil}$, the runaway pile up is significantly suppressed. However, as we will show later, if we also add Diff-BKR (the back-reaction for the radial diffusion), even with Eq. (\ref{eq_silicate_H}), the runaway pile-up again occurs for reasonable values of $F_{\rm p/g}$.\\

In the rest of this paper, we include Drift-BKR of both pebbles and silicate grains that slow down their radial velocities as they pile up. \\
%

\subsubsection{Typical results of a steady state}
\label{sec_steady_state}
\begin{figure*}
\resizebox{\hsize}{!}{ \includegraphics{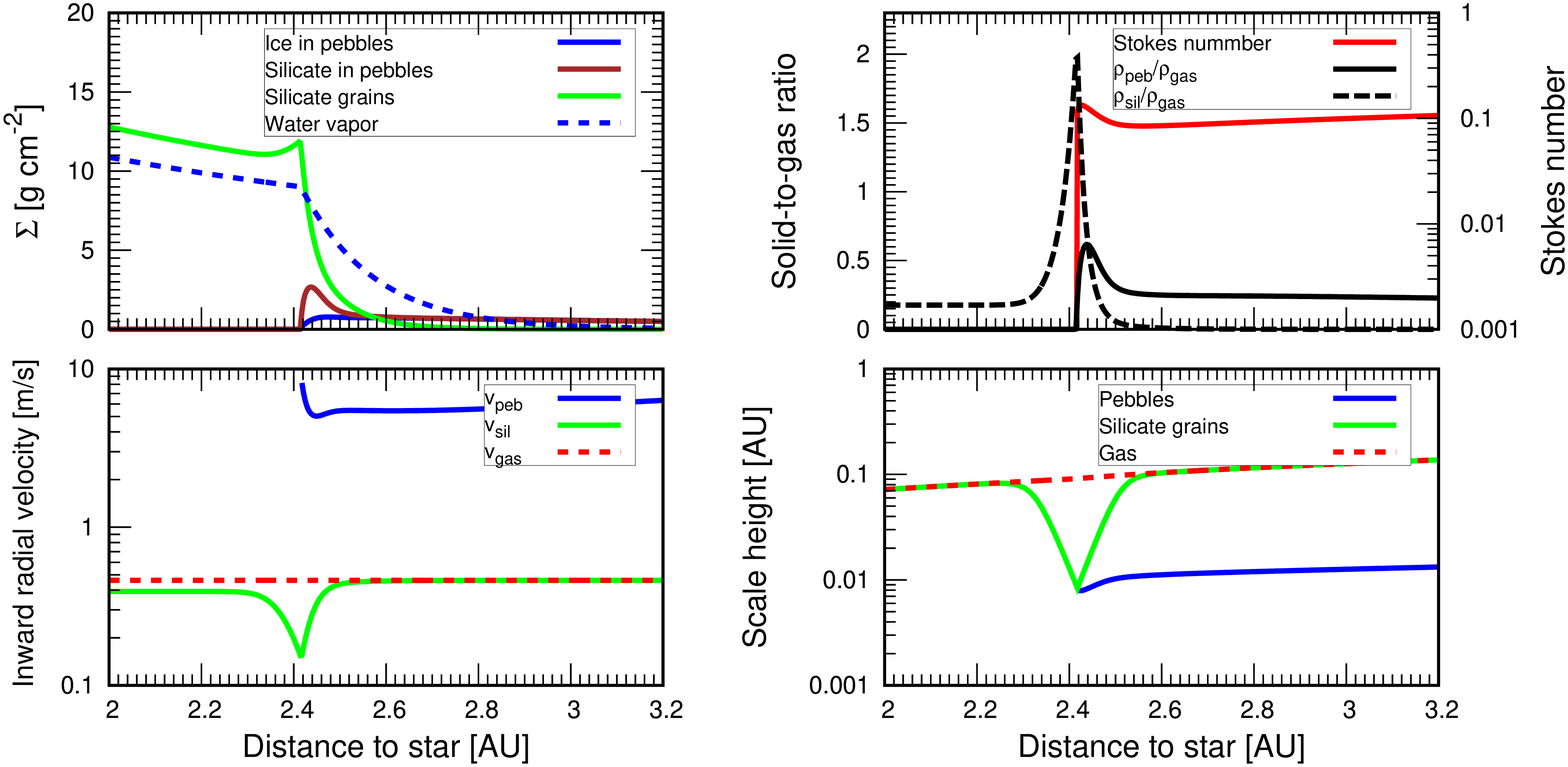}}
\caption{Typical steady state in the case of $\alpha_{\rm acc}=1\times10^{-2}$, $\alpha_{\rm tur}=1\times10^{-3}$, $F_{\rm p/g}=0.3$ and Drift-BKR is included ($K=0$). Left top panel shows surface densities of solids and vapor. Right top panel shows midplane solid-to-gas ratios of pebbles (solid line) and silicate grains (dashed line) as well as the Stokes number of pebbles by red solid line. Left bottom panel shows radial velocities of solids and vapor. Right bottom panel shows scale heights of solids and vapor. In each panel (except top right panel), blue and brown solid lines represent those of ice and silicate components in pebbles, respectively. Green lines represent those of silicate grains. Red and blue dashed lines represent those of the background gas and water vapor, respectively.}
\label{typical_results}
\end{figure*}

Figure \ref{typical_results} shows results of a typical example of our simulations when the system reaches a steady state (a $K=0$ case with $F_{\rm p/g}=0.3$, $\alpha_{\rm tur}=10^{-3}$, $\alpha_{\rm acc}=10^{-2}$ and $\dot{M}_{\rm gas}=10^{-8}M_{\rm s}$/year). The steady state is reached by $3-4 \times 10^{5}$ years. Silicate grains released by sublimation of icy pebbles pile up just inside the water snow line (due to "traffic jam" effect: silicate grains are well coupled to the gas and they drift much slower than pebbles). In our model, silicate grains diffuse vertically from pebbles at the snow line and their scale heights gradually become larger to reach that of the gas as the distance to the snow line becomes larger (Figure \ref{typical_results} right bottom panel; see also Eq. (\ref{eq_silicate_H})). Thus, the midplane solid-to-gas ratio of silicate grains has a peak just inside the snow line where silicate grains' scale-height is the smallest (Figure \ref{typical_results} right top and bottom panels, respectively). Since the radial drift velocity of silicate grains is much smaller than that of pebbles, the surface density of silicate grains become much larger than that of pebbles (Figure \ref{typical_results} left top panel). Just outside the snow line, the water vapor diffuses outward and recondense ("cold finger" effect: a part of the water vapor inside the snow line is returned outside of the snow line through its diffusion and recondenses outside the snow line) and silicate grains also diffuse outward and they stick to icy pebbles (Eq. (\ref{eq_silicate_stick})), resulting in an enhancement of pebble pile-up \citep[see also][]{Dra17,Sch17}. As pebbles and silicate grains pile up outside and inside the snow line, respectively, their drift velocities decrease due to Drift-BKR of pebbles and silicate grains around the snow line (Figure \ref{typical_results} left bottom panel).\\

%

\subsubsection{Dependence on the turbulent viscosity}
\begin{figure*}
\resizebox{\hsize}{!}{ \includegraphics{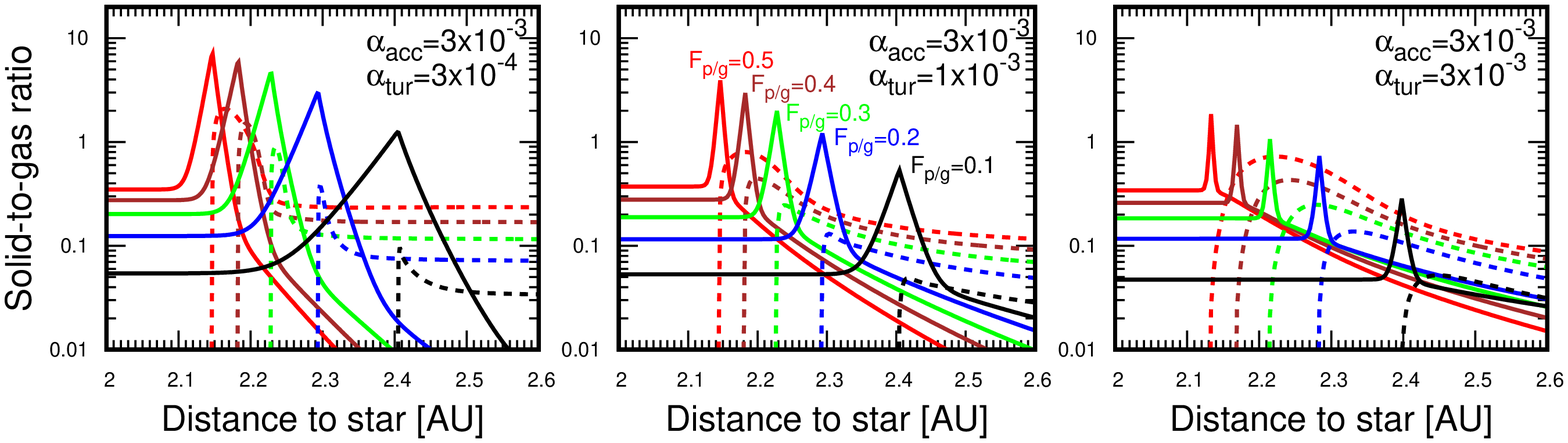}}
\caption{Midplane solid-to-gas ratio of silicate grains (solid lines) and pebbles (dashed lines) at different diffusion strength $\alpha_{\rm tur}$ and pebble flux $F_{\rm p/g}$ at effective alpha of $\alpha_{\rm acc}=3\times10^{-3}$. The left, middle, and right panels show the results for $\alpha_{\rm tur} = 3 \times 10^{-4}, 10^{-3}$ and $3 \times 10^{-3}$, respectively. Red, brown, green, blue and black lines represent cases of $F_{\rm p/g}=0.5, 0.4, 0.3, 0.2$ and $0.1$, respectively. Silicate grains have the peaks of their solid-to-gas ratios just inside the snow line. This is because the silicate grains initially have the same scale height with that of pebbles and they gradually diffuse vertically to reach the gas scale height (see Eq. (\ref{eq_silicate_H}))}
\label{fig_K0_cases}
\end{figure*}

Figure \ref{fig_K0_cases} shows the solid-to-gas ratios in steady states for different values of $\alpha_{\rm tur}$ with different scaled pebble flux $F_{\rm p/g}$. The value of $\alpha_{\rm acc}$ is fixed to $\alpha_{\rm acc}=3 \times 10^{-3}$ in all cases. The location of peaks moves inward and the peak height increases as $F_{\rm p/g}$ increases. Because $\Sigma_{\rm vap}$ and accordingly $P_{\rm vap}$ increase with $F_{\rm p/g}$, the snow line location that is determined by $P_{\rm vap}=P_{\rm eq}$ shifts inward. As $F_{\rm p/g}$ becomes larger, more solids exist in the gas disk and then the peaks become larger. The peaks of silicate grains inside the snow line increase with the decrease in $\alpha_{\rm tur}$ for the same $F_{\rm p/g}$. Because $\alpha_{\rm tur}$ represents the strength on vertical mixing and the scale height of silicate grains is smaller for smaller $\alpha_{\rm tur}$, the concentration of silicate grains is higher for smaller $\alpha_{\rm tur}$. Peaks of pebbles also show the same dependence as silicate grains. As $\alpha_{\rm tur}$ becomes larger to be closer to $\alpha_{\rm acc}$, the peaks of silicate grains become comparable to those of pebbles outside the snow line, because diffusions of silicate grains and water vapor from inside of the snow line to outside of the snow line become efficient. It means that GI of silicate grains inside the snow line becomes more important than SI of icy pebbles outside the snow line as $\alpha_{\rm tur}/\alpha_{\rm acc}$ decreases.\\
%

\subsubsection{Composition of pile-up-solids}
\begin{figure*}
\resizebox{\hsize}{!}{  \includegraphics{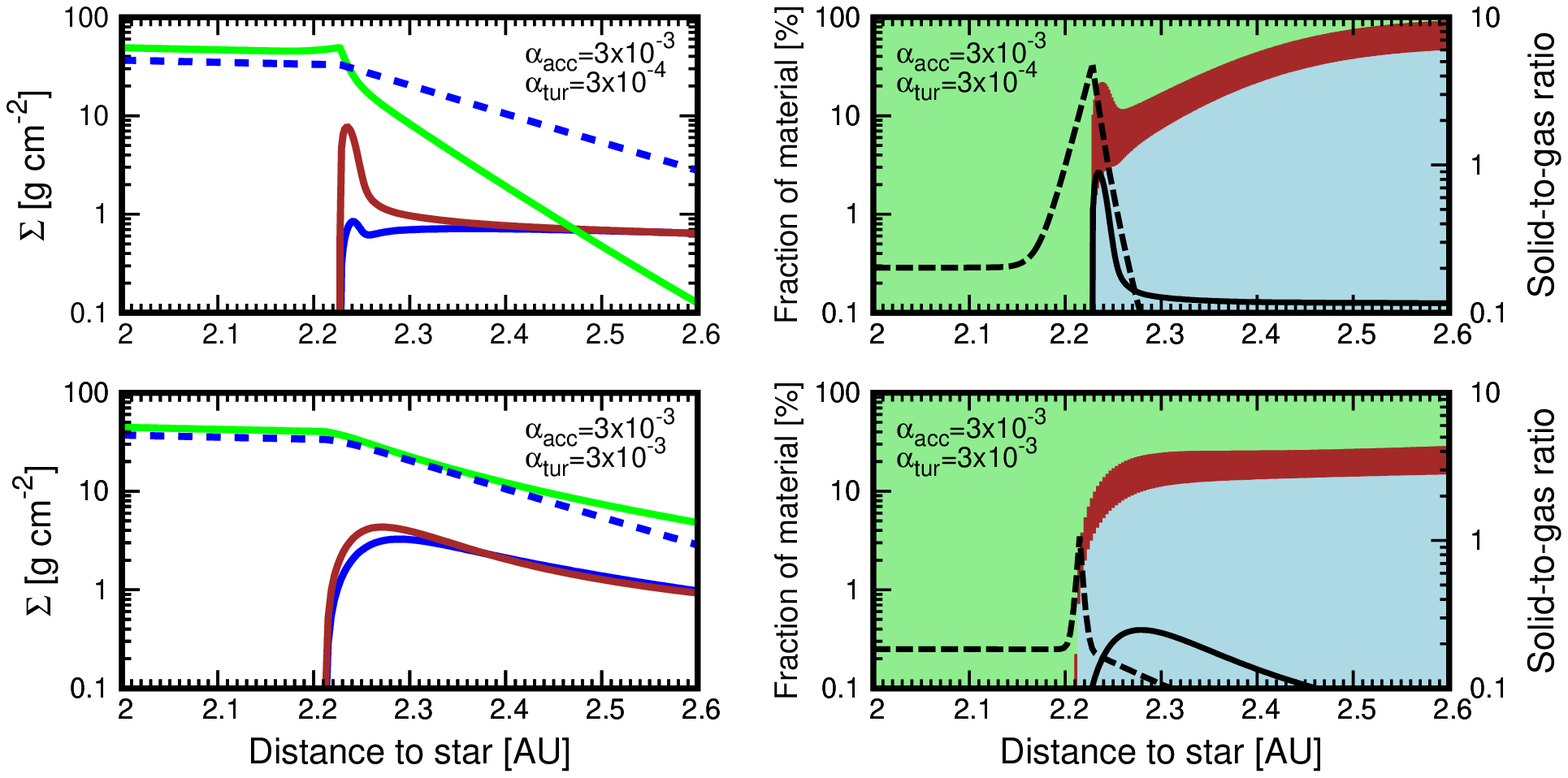}}
\caption{Surface density profile (left) and fraction of different materials (right) in two cases of turbulent strength, $\alpha_{\rm tur}=3\times10^{-4}$ (top) and $\alpha_{\rm tur}=3\times10^{-3}$ (bottom) with $\alpha_{\rm acc}=3\times10^{-3}$ and $F_{\rm p/g}=0.3$ ($K=0$). In the left panels, green, brown and blue solid lines represent surface densities of silicate grains, silicate in a pebble and ice in a pebble, respectively. Blue dashed lines in the left panels represent vapor surface density. In the right panel, green, brown and blue colors represent a fraction of silicate grains, silicate in a pebble and ice in a pebble, respectively. In the right panel, dashed and solid lines represent the solid-to-gas ratio of silicate grains and pebbles, respectively.}
\label{fig_composition}
\end{figure*}

In Figure \ref{fig_composition}, we show the composition of solids at the different radial distances to the star. Silicate grains dominate inside the snow line. On the other hand, silicate grains and pebbles coexist outside the snow line. Just inside the snow line, silicate grains pile up and the solid-to-gas ratio at the midplane can be much larger than unity. This is because, around the snow line, silicate grains have small scale height as that of pebbles (Figure \ref{typical_results} right bottom panel or Eq. (\ref{eq_silicate_H})). This indicates that such a location is a favorable place to form rocky planetesimals by gravitational instability \citep[see also][]{Ida16} but not by streaming instability because of the small Stokes number of silicate grains (Figure \ref{typical_results}, upper right panel). In contrast, outside of the snow line is a favorable place to form ice-rich (mixture of rock and ice) planetesimals via streaming instability because a pebble has the Stokes number large enough to operate SI \citep[see also][]{Dra17,Sch17}.\\

The fraction of silicate-to-ice strongly depends on $\alpha_{\rm tur}$ (Figure \ref{fig_composition}). Figure \ref{fig_composition} shows cases of two different turbulent viscosities, $\alpha_{\rm tur}=3\times10^{-3}$ and $3\times10^{-4}$ with $\alpha_{\rm acc}=3\times10^{-3}$. As discussed above, with smaller $\alpha_{\rm tur}$, the pile-up of silicate grains is more enhanced just inside the snow line (see also Figure \ref{fig_K0_cases}), potentially forming rocky planetesimals via gravitational instability. Since the pile-up of the silicate grains is locally enhanced just inside the snow line, its outward diffusion toward the snow line is also enhanced compared to that of water vapor and thus silicate grains efficiently stick to icy pebbles outside of the snow line $-$ the sticking efficiency is further enhanced because silicate grains around the snow line has small scale height ($R_{\rm s} \propto 1/H_{\rm sil}$ and $H_{\rm sil}$ becomes smaller for smaller $\alpha_{\rm tur}$). Therefore, just outside the snow line, the fraction of silicate grains inside a pebble becomes larger with $\alpha_{\rm tur}=3\times10^{-4}$ compared to the case of $\alpha_{\rm tur}=3\times10^{-3}$. This suggests that water-poor (relatively dry) planetesimals may also form via streaming instability even outside the water snow line when the turbulent viscosity is small $-$ the fraction of silicate material becomes larger than that of water ice as outward diffusion flux of silicate grains becomes larger than that of water vapor as the turbulent viscosity becomes smaller.\\

%
\subsubsection{Dependence on the effective accretion viscosity}
\begin{figure*}
\resizebox{\hsize}{!}{  \includegraphics{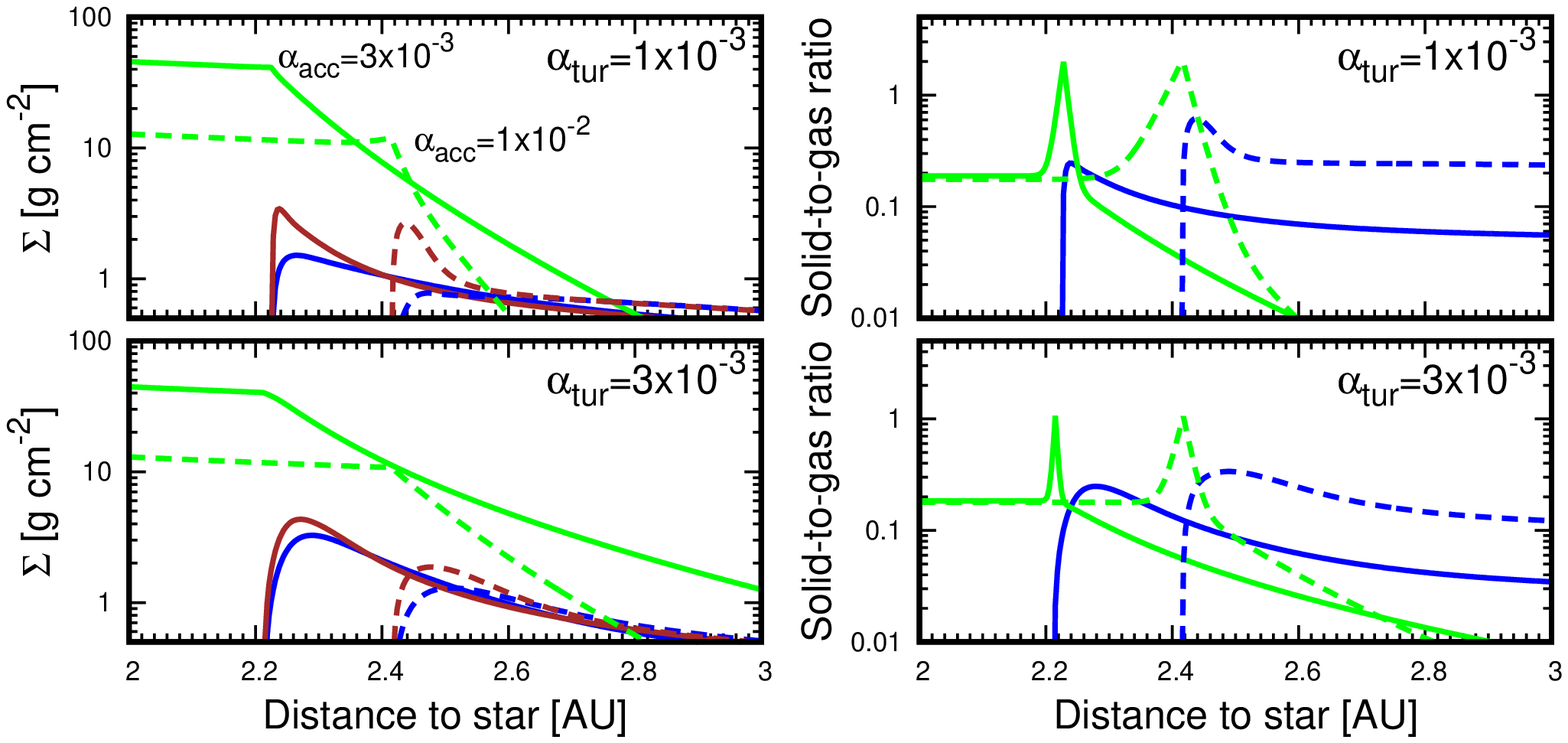}}
\caption{Surface density (left) and midplane solid-to-gas ratio (right) for the case of $F_{\rm p/g}=0.3$, $\dot{M}_{\rm gas}=10^{-8} M_{\rm sun}$/year and $K=0$. Top panels show the cases of the turbulent viscosity of $\alpha_{\rm tur}=1\times10^{-3}$. Bottom panels show the case of $\alpha_{\rm tur}=3\times10^{-3}$.  In each panel, solid and dashed lines represent the cases of the effective viscosity of $\alpha_{\rm acc}=3\times10^{-3}$ and $\alpha_{\rm acc}=1\times10^{-2}$, respectively. In the left panel, blue, brown and green lines represent those of ice in pebbles, silicate in pebbles and silicate grains, respectively. In the right panel, blue and green lines represent those of pebbles and silicate grains, respectively.}
\label{fig_acc_dependence}
\end{figure*}

In this subsection, we study the dependence on the effective viscosity $\alpha_{\rm acc}$ that regulates the accretion of the gas to the central star. Figure \ref{fig_acc_dependence} shows results of our simulations ($\dot{M}_{\rm gas}=10^{-8} M_{\rm sun}$/year and $K=0$). As the effective viscosity decreases, surface densities of both the gas and solids increase ($\Sigma_{\rm gas} \propto 1/\alpha_{\rm acc}$) and the snow line shifts inward due to the effect of water vapor pressure.\\
 
As explained in the previous section, the pile-up of silicate grains inside the snow line is more significant for smaller turbulent viscosity $\alpha_{\rm tur}$. The resultant pebbles contain more silicate component for smaller turbulent viscosity than ice $-$ potentially forming silicate-rich planetesimals even outside the snow line. In contrast, as the turbulent viscosity (described by $\alpha_{\rm tur}$) becomes close to the value of the accretion viscosity (described by $\alpha_{\rm acc}$), the composition becomes almost an equal mixture of silicate and ice (see also results of \cite{Sch17} where they assume $\alpha_{\rm acc}=\alpha_{\rm tur}$).\\

The peak values of surface densities of pebbles and silicate grains become larger and the Stokes number of a pebble at the snow line becomes smaller for smaller accretion viscosity ($\alpha_{\rm acc}$). However, the solid-to-gas ratio of pebbles becomes smaller for smaller $\alpha_{\rm acc}$ because the gas surface density becomes larger and $\rho_{\rm peb}/\rho_{\rm gas} \propto (\Sigma_{\rm peb}/\Sigma_{\rm gas}) \times ({\rm St}/\alpha_{\rm tur}+1)^{-1/2}$. \\
%

\subsubsection{Dependence on the accretion rate}
\label{sec_dep_accretion_rate}
\begin{figure*}
\resizebox{\hsize}{!}{ \includegraphics{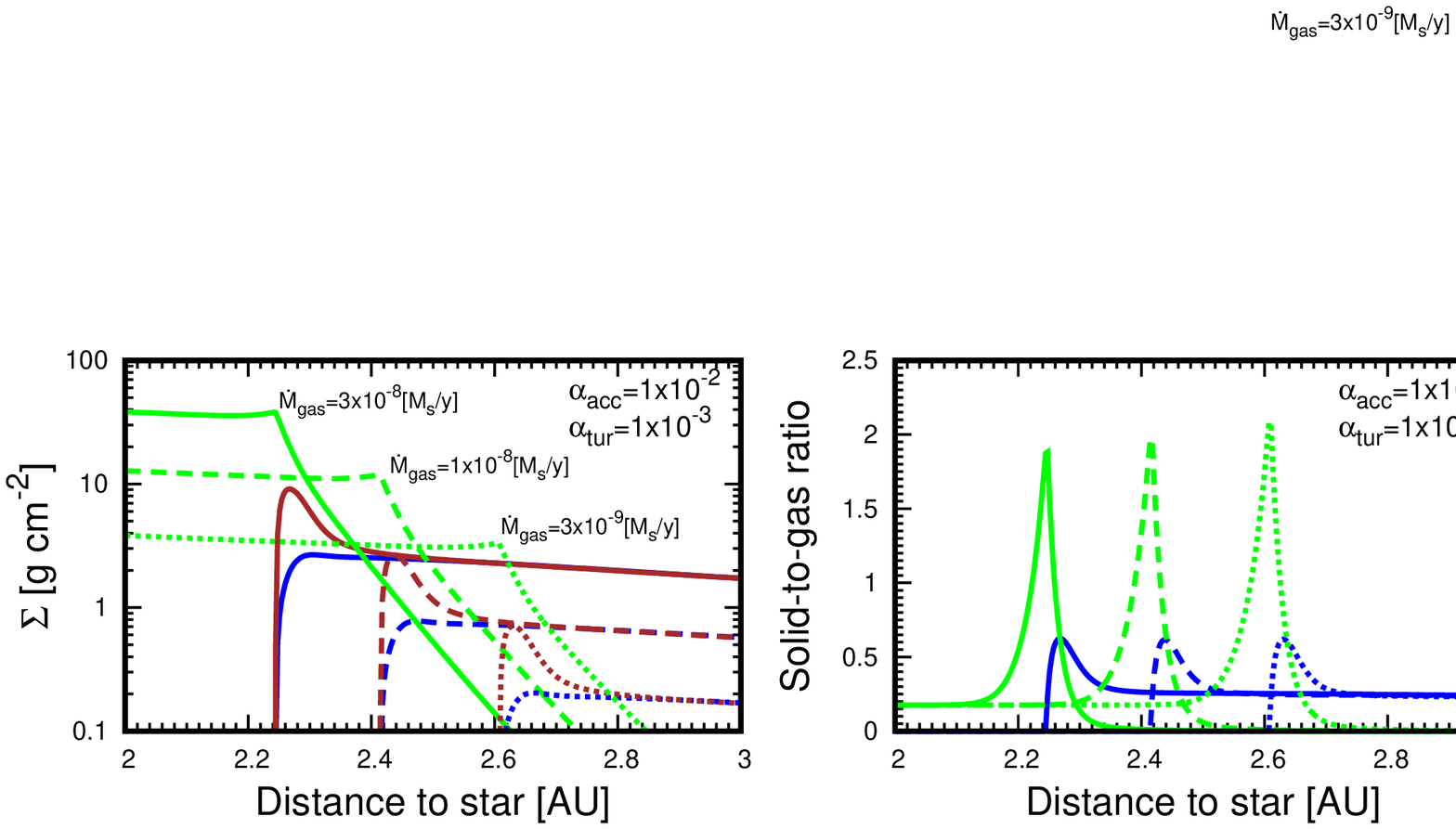}}
\caption{Surface density (left) and midplane solid-to-gas ratio (right) for the case of $\alpha_{\rm acc}=10^{-2}$, $\alpha_{\rm tur}=10^{-3}$, $F_{\rm p/g}=0.3$ and $K=0$. In each panel, solid, dashed and dotted lines represent the cases of different accretion rate of the gas as $\dot{M}_{\rm gas}=3\times10^{-8} M_{\rm sun}$/year, $\dot{M}_{\rm gas}=1\times10^{-8} M_{\rm sun}$/year and $\dot{M}_{\rm gas}=3\times10^{-9} M_{\rm sun}$/year, respectively. In the left panel, blue, brown and green lines represent those of ice in pebbles, silicate in pebbles and silicate grains, respectively. In the right panel, blue and green lines represent those of pebbles and silicate grains, respectively.}
\label{fig_Mdot_dependence}
\end{figure*}

In this subsection, we study the dependence on the accretion rate of the gas. As the accretion rate increases, surface densities of the gas, pebbles, and silicate grains increase. A higher accretion rate increases vapor surface density and thus the snow line shifts inward due to its pressure effect (Figure \ref{fig_Mdot_dependence}). Note that if disk temperature is determined by the viscous heating, the increase in the accretion rate enhances the viscous heating and the snow line may shift outward.\\

Even though the surface density of pebbles and silicate grains increases for larger mass accretion rate, the midplane solid-to-gas ratio does not change significantly because the gas surface density becomes also larger for larger accretion rate (Figure \ref{fig_Mdot_dependence}).\\
%

\subsection{Case including the diffusion back-reactions (K=1 and 2)}
\label{sec_K1K2}
Here, we discuss the effect of Diff-BKR ($K = 1$ and $2$) that weakens diffusion of solids as the pile-up of solid proceeds (Eqs. (\ref{eq_D_pebble}) and (\ref{eq_D_silicate})). We study the dependence on different diffusion viscosity between $\alpha_{\rm tur}=1\times10^{-3}$ and $1\times10^{-2}$.\\

\subsubsection{Runaway pile-ups of solids inside and outside the snow line}
\begin{figure*}
\resizebox{\hsize}{!}{ \includegraphics{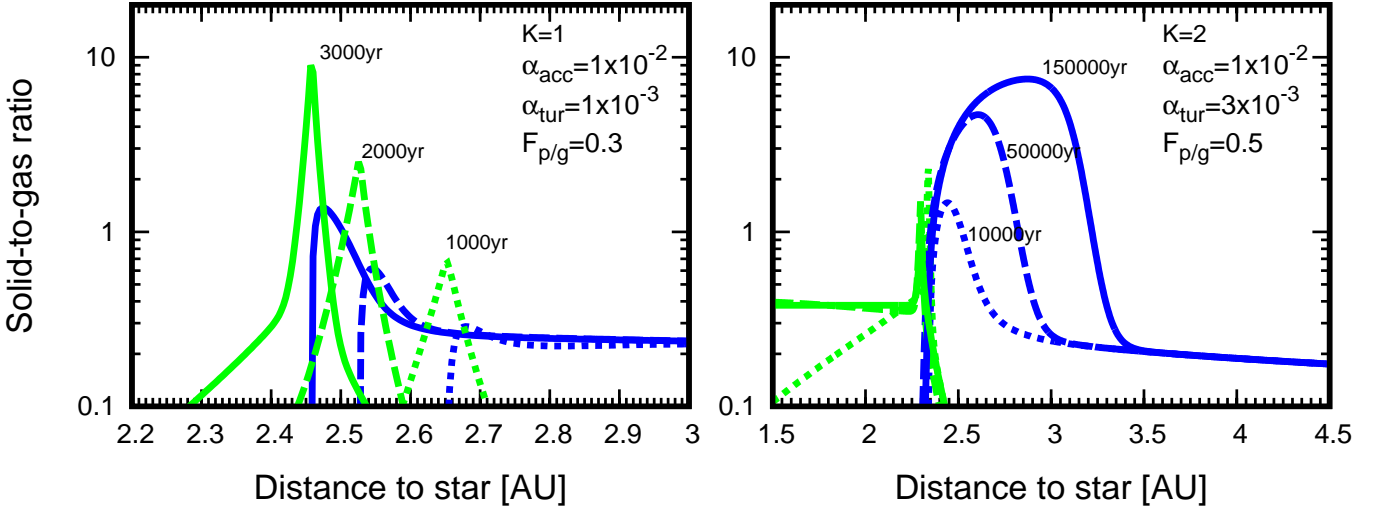}}
\caption{Runaway evolution of midplane solid-to-gas ratios in the cases of Sil-RPU (left) and Ice-RPU (right). Blue and green lines represent midplane solid-to-gas ratio of pebbles and silicate grains, respectively. Time proceeds as dotted, dashed and solid lines. Left panel is the case of $\alpha_{\rm tur}=1\times10^{-3}$, $F_{\rm p/g}=0.3$, $\alpha_{\rm acc}=1\times10^{-2}$ and $K=1$. Right panel is the case of $\alpha_{\rm tur}=3\times10^{-3}$, $F_{\rm p/g}=0.5$, $\alpha_{\rm acc}=1\times10^{-2}$ and $K=2$.}
\label{fig_RPs}
\end{figure*}

When Diff-BKR is included ($K \neq 0$), the diffusion coefficient becomes progressively smaller as the pile-up of silicate grains or icy pebbles proceeds (see Eqs. (\ref{eq_D_pebble}) and (\ref{eq_D_silicate})). When the diffusion becomes weaker than the inflow of silicate grains or pebbles to the region inside or outside the snow line, a runaway pile-up occurs and the system never reaches a steady state (Figure \ref{fig_RPs}).\\

We found two different runaway pile-ups potentially occur (Figure \ref{fig_RPs}) $-$ (1) A runaway pile-up of silicate grains inside the snow line (hereafter "Sil-RPU") and (2) a runaway pile-up of pebbles that contains both silicate and ice outside the snow line (hereafter "Ice-RPU"). As Figure \ref{fig_RPs} shows, Sil-RPU is favored in the case of $\alpha_{\rm tur}/\alpha_{\rm acc} \ll 1$, while Ice-RPU is favored in the case of $\alpha_{\rm tur}/\alpha_{\rm acc} \sim 1$. In the former case, water vapor and released silicate grains do not efficiently diffuse outside the snow line and thus silicate grains continuously pile up inside the snow line, resulting in Sil-RPU. In the latter case, water vapor efficiently diffuses outside the snow line, resulting in Ice-RPU $-$ the diffusion of water vapor from inside to outside the snow line enhances the pile-up of icy pebbles and the diffusivity of water vapor is not affected by the pile-up of solids (Eq. (\ref{eq_D_gas})). In contrast, the diffusivity of icy pebbles becomes progressively smaller as the pile-up proceeds by Diff-BKR (Eq. (\ref{eq_D_pebble})), resulting in Ice-RPU.\\

\subsubsection{Disk parameter range for runaway pile-ups}
\begin{figure*}
\resizebox{\hsize}{!}{ \includegraphics{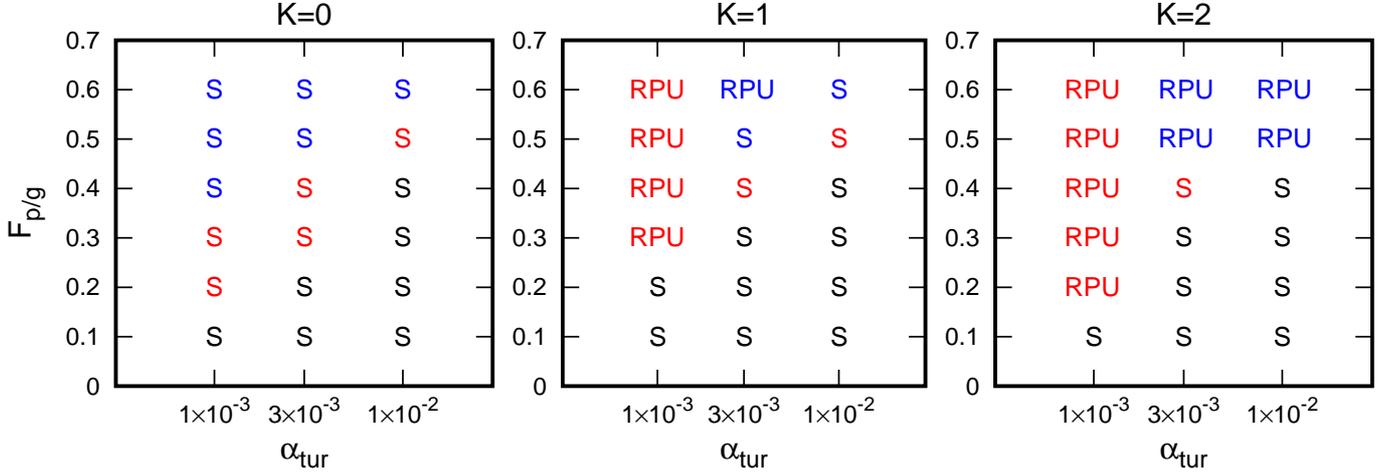}}
\caption{Parameter maps where either silicate grains' runaway pile-ups (labeled "RPU" in red color), icy pebble runaway pile-ups (labeled "RPU" in blue color) or steady state without runaway pile-ups (labeled "S" in black color) occurs for different $\alpha_{\rm tur}$ and $F_{\rm p/g}$ ($\alpha_{\rm acc}=10^{-2}$). If the midplane dust-to-gas ratio of silicate grains is larger than unity for the case of "S", the label is highlighted in red color. If the midplane dust-to-gas ratios of pebbles as well as those of silicate grains are larger than unity for the case of "S", the label is highlighted in blue color. Left, middle and right panels show the cases of $K=0$, $K=1$ and $K=2$, respectively.}
\label{fig_K1K2}
\end{figure*}

In Figure \ref{fig_K1K2}, the parameter regions for Sil-RPU and Ice-RPU are summarized for $K$ = 0, 1, and 2. When $K=0$, the system reaches the steady state for $\alpha_{\rm tur} \geq 1\times 10^{-3}$ and $\alpha_{\rm acc}=1\times10^{-2}$. As $\alpha_{\rm tur}/\alpha_{\rm acc}$ becomes smaller, the diffusion becomes less efficient and the scale height of solids becomes smaller (see Eqs. (\ref{eq_H_pebble}) and (\ref{eq_silicate_H})). This leads to more efficient pile-up of silicate grains inside the snow line and the midplane solid-to-gas ratio easily grows much larger than unity even for small pebble flux (e.g. $F_{\rm p/g}=0.2$ and $\alpha_{\rm tur}=1\times10^{-3}$; see Figure \ref{fig_K1K2} left panel). As the pebble-to-gas mass flux $F_{\rm p/g}$  becomes larger, not only silicate grains inside the snow line but also pebbles outside the snow line more efficiently pile up and the solid-to-gas ratio of pebbles becomes larger than unity (the parameter regions are marked by red and blue "S" in Figure \ref{fig_K1K2}; see also Figure \ref{fig_K0_cases}). Because pebbles have the Stokes number large enough for SI to operate, SI is expected for the parameters marked by blue "S".  As $\alpha_{\rm tur}$ increases up to $\sim \alpha_{\rm acc}$, the diffusion becomes efficient and the pile-ups of solids are efficiently smoothed and the midplane solid-to-gas ratios are reduced. In other words, in order to reach higher concentrations of solids at the midplane, larger $F_{\rm p/g}$ is required for higher value of $\alpha_{\rm tur}$. Our results are consistent with those in \cite{Sch17} where they assumed $K=0$ and $\alpha_{\rm tur}=\alpha_{\rm acc}$ and where they reported outside the snow line is a favorable place for SI. However, our results are different from those in \cite{Sch17} where they reported that the midplane solid-to-gas ratio of silicate grains inside the snow line never reach above unity. This is because they assumed that the silicate grains are instantaneously mixed with the background gas ($H_{\rm sil} = H_{\rm gas}$) and neglected Drift-BKR of silicate grains. In contrast, we take into account Drift-BKR of the silicate grains and consider that the silicate grains released from pebbles have scale height that is initially the same as those of pebbles at the snow line and diffuse vertically up to gas scale height (see Eq. (\ref{eq_silicate_H})).\\

When $K=1$ or $K=2$, as the pile-ups of solids proceed, the diffusion of solids becomes further weakened (Eqs. (\ref{eq_D_pebble}) and (\ref{eq_D_silicate})) and the pile-ups of solids occur in a runaway fashion for sufficiently large $F_{\rm p/g}$ (see Figure \ref{fig_K1K2}). As in the case of $K$ = 0, Sil-RPU is more favored than Ice-RPU for $\alpha_{\rm tur}/\alpha_{\rm acc} \ll 1$ in the cases of $K$ = 1 and 2. In the limit of inefficient diffusion, the situation is similar to what \cite{Ida16} assumed, that is, our simulations of $K \neq 0$ cases somewhat mimic the situation (no diffusion) of \cite{Ida16} because our modeled diffusivity progressively becomes smaller by Diff-BKR as the pile-up proceeds (Eqs. (\ref{eq_D_pebble}) and (\ref{eq_D_silicate})). When $\alpha_{\rm tur}/\alpha_{\rm acc} \sim 1$, the pile-up of pebbles is enhanced and thus Ice-RPU favorably occurs. These runaway pile-ups more easily occur as $K$ becomes larger (Figure \ref{fig_K1K2}) because the degree of diminishment in the diffusion coefficient is enhanced at the same solid-to-gas ratio (see Eqs. (\ref{eq_D_pebble}) and (\ref{eq_D_silicate})). In the Sil-RPU/Ice-RPU cases, the numerical calculations show that the midplane solid-to-gas ratio grows to infinity without reaching a steady state, which strongly suggests that rocky or icy planetesimals will be formed.\\

We note that one-dimensional code may not precisely predict the parameter boundary where Sil-RPU and Ice-RPU occurs because there is no steady state solution $-$ The boundary in Figure \ref{fig_K1K2} can change as changing the simulation timestep or grid size in a way, for example, that runs with smaller timesteps require larger pebble flux ($F_{\rm p/g}$) for the peaky Sil-RPU to occur. This would be due to the fact that smaller timesteps are accompanied by larger numerical viscosity at finite-difference methods. In other words, as the pile-up of solids proceeds, the local diffusion viscosity decreases due to our modeled diffusion back-reaction (see Eqs. (\ref{eq_D_pebble}) and (\ref{eq_D_silicate})). At some point, artificial numerical diffusion overwhelms the local physical diffusion and then the pile-up is regulated by this artificial numerical viscosity. This is the numerical limitation of one-dimensional codes and thus further studies on Sil-RPU and Ice-RPU are required. In this work, we use small timesteps enough to solve the motion of dust and gas $-$ the Courant Parameter is well below the unity $-$ and thus Figure \ref{fig_K1K2} may predict the parameter boundary at larger $F_{\rm p/g}$ than the actual value. We will leave this issue for future studies.\\
%

\section{Discussion}
\label{sec_discussion}
\subsection{Implications for planet formation}
\label{sec_planet_formation}

In this work, we show that when $\alpha_{\rm tur}/\alpha_{\rm acc} \ll 1$, the runaway pile-up of silicate grains inside the snow line (Sil-RPU) preferentially occurs and formation of rocky planetesimals by GI is favored even for relatively small ($F_{\rm p/g} \sim 0.2-0.3$) pebble mass flux (Sil-RPU in Figure \ref{fig_K1K2}). When $\alpha_{\rm tur}/\alpha_{\rm acc} \sim 1$, the runaway pile-up of icy pebbles outside the snow line (Ice-RPU) occurs and formation of icy planetesimals by SI is favored for sufficient pebble mass flux (Ice-RPU in Figure \ref{fig_K1K2}).\\

During the evolution of protoplanetary disks, the snow line migrates inward as the mass accretion rate decreases in time \citep[e.g.][]{Oka11}. The disk inner region may have a $dead$ $zone$ near the disk midplane where turbulence is weak ($\alpha_{\rm tur}$ is small) \citep[e.g.][]{Gam96}. The radial boundary between the active zone in the outer disk region ($\alpha_{\rm tur} \sim \alpha_{\rm acc}$) and the dead zone in the inner disk region ($\alpha_{\rm tur} \ll \alpha_{\rm acc}$) might be not sharp \cite[e.g.][]{Bai16,Mor17}. Therefore, it may be expected that during the early phase of the disk evolution when the disk accretion rates is high, the snow line is located in the relatively outer region of the disk where the disk is active ($\alpha_{\rm tur} \sim \alpha_{\rm acc}$; e.g. several AU around the solar-type star) and the pile-up of icy pebbles may efficiently occur outside the snow line to form icy planetesimals by GI or SI with sufficiently high pebble mass flux (see Figure \ref{fig_K1K2}). The icy planetesimals can produce icy cores of giant planets.  In the later phase of the disk evolution, the snow line migrates inward (e.g. $\sim$ 1 AU) where the disk midplane is dead ($\alpha_{\rm tur} \ll \alpha_{\rm acc}$) and the runaway pile-up of silicate grains may efficiently occur even for relatively small pebble mass flux (see Figure \ref{fig_K1K2}), resulting in the formation of rocky planetesimals by GI. The results of Figure \ref{fig_K1K2} may also suggest that in the intermediate phase when $\alpha_{\rm tur}$ is an intermediate value ($\alpha_{\rm tur} < \alpha_{\rm acc}$ but not $\alpha_{\rm tur} \ll \alpha_{\rm acc}$) and the pebble mass flux is not enough for SI to operate, the snow line could be around the asteroid region (e.g. $\sim$ 2 AU) and thus an efficient planetesimal formation either by SI or GI might not be established in the asteroid region. This scenario potentially provides rocky planetesimals confined near $1$ AU, from which the terrestrial planet configuration in our Solar system is naturally reproduced \citep[e.g.][]{Han09}.\\

\subsection{Potential caveats in our study}
\label{sec_caveats}
In this study, we have several assumptions and we will comment on them as follows.\\

The scale height of silicate grains: we assume that silicate grains at the snow line initially have the same scale height as that of pebbles at the snow line. This is partly correct because silicate grains are released from pebbles. But the sublimation takes place not only at the snow line but it progressively occurs as pebbles approach the snow line. Such early-released grains may be stirred up to have a larger scale height than that estimated by Eq. (\ref{eq_silicate_H}).  However, since the sublimation rate exponentially depends on the disk temperature, and accordingly on the distance from the star, the fraction of the early-released grains is negligible.\\

Back-reaction of the vertical stirring: in this paper, we assume that the vertical diffusion of solids is unaffected by pile-ups of solids and thus their scale heights are not a function of solid-to-gas ratio $Z$, that is, we assume that $H_{\rm peb}$ and $H_{\rm sil}$ are independent of $Z$ (Eqs. (\ref{eq_H_pebble}) and (\ref{eq_silicate_H})). However, in reality, in the same way as the radial diffusivity of solids is potentially weakened by pile-ups of solids as back-reaction (Diff-BKR), the vertical diffusion can be also weakened. In this case, the scale heights of solids progressively become smaller as pile-up proceeds. This would lead to further enhancement of the midplane solid-to-gas ratio, resulting in a more favorable condition of planetesimal formation. These effects should be studied in the future works.\\

Growth of particles: we ignore the coagulation of silicate grains to larger silicate particles, because simulations and experiments strongly suggest that a bouncing or fragmentation barrier is severe for collisions between silicate grains \citep[e.g.][]{Blu00,Zso11,Wad11} $-$ we only consider pebble growth by sticking of silicate grains and condensation of water vapor onto pebbles outside of the snow line. If coagulation among silicate grains is efficient, they become larger and thus the Stokes number increases. Then, their radial drifts increase as well as their outward diffusion beyond the snow line becomes less efficient. These effects may reduce the surface density of silicate grains inside the snow line or/and that of pebbles outside the snow line, potentially leading to a decrease of their solid-to-gas ratio. However, at the same time, the scale height of silicate grains becomes smaller as they grow and this leads to larger solid-to-gas ratio. This complex process is left for future work. In this paper, we try to use a relatively simple model, in order to highlight the new finding of how rocky or icy planetesimal formation near the snow line depends on the disk parameters.\\

Diffusivity of solids: in this paper, we assume that the solid diffusivity drops with increasing its solid-to-gas ratio at midplane, $Z$. We assume that the diffusivity decreases as $(1+Z)^{-K}$ ($K=1$ or $2$; Eqs. (\ref{eq_D_pebble}) and (\ref{eq_D_silicate})). However, it is not understood that how the pile-up of solids affects their diffusivity and thus the dependence on $Z$ is not known. We will leave this subject in future work. Note that, however, in the limit of $Z \to 0$, the back-reaction vanishes, and the efficiency of radial diffusion is independent of solid surface density. In the other limit of $Z \to \infty$, the radial diffusivity goes to zero, because gas density vanishes. In reality, angular momentum exchanges by collision and gravitational scattering between grains/pebbles should result in radial diffusion. However, it would be negligible and $Z \to \infty$ already implies the formation of planetesimals. \\

Motion of the gas: in this work, we considered back-reactions of solids to the gas that affect the motion of solids $-$ radial back-reactions and diffusion back-reactions (Eqs. (\ref{eq_BR_solid1}), (\ref{eq_BR_solid2}), (\ref{eq_D_pebble}) and (\ref{eq_D_silicate})) $-$ but we neglected the changes in the motion of the gas. However, as the solids pile up, the modulation in the disk gas structure due to the back-reaction would not be negligible. Inclusion of the modulation of the disk gas structure will be left for future work.\\

\section{Summary}
\label{sec_summary}
It is challenging to form tens-to-hundreds kilometer-sized planetesimals from micron-sized dust growing through all the intermediate sizes due to growth barrier and radial drift barrier. Many authors focused on inside/outside the water snow line as a special location where solids (silicates, ices) efficiently pile up. Streaming instability outside of the snow line \citep{Arm16,Dra17,Sch17} or by gravitational instability inside of the snow line \citep{Ida16} are possible mechanisms to form planetesimals directly from small particles without growing through all the intermediate sizes.\\

Through one-dimensional simulation for sublimation/re-condensation and migration of icy pebbles, \cite{Sch17} investigated the possibility of streaming instability (SI) of the piled-up icy pebbles just outside the water snow line. While they took into account the back-reaction of the pebbles in radial drift velocity (Drift-BKR; Eqs. (\ref{eq_BR_solid1}) and (\ref{eq_BR_solid2})), they did not include Drift-BKR of the silicate grains released from the sublimating icy pebbles. They also assumed that the grains are instantaneously mixed with the background gas and the scale height of the grains is always the same as that of the background gas (that is Eq. (\ref{eq_H_gas})) which is much larger than that of pebbles (Eq. (\ref{eq_H_pebble})). As a result, they did not find a runaway pile-up of the released silicate grains inside the snow line (Sil-RPU) that \cite{Ida16} proposed. Sil-RPU would result in formation of rocky planetesimals by gravitational instability. We note that \cite{Ida16} focused on Sil-RPU and used a simple analytical model in which the radial diffusion of water vapor and that of the grains are neglected, while they considered the Drift-BKR of the silicate grains that \cite{Sch17} neglected. \\

In this study, following the 1D simulation of diffusion/advection/sublimation/recondensation of silicate/ice/gas used by \cite{Sch17}, we have further studied the pile-up of both icy pebbles and the released silicate grains outside/inside the snow line, taking into account Drift-BKR of the grains as well as that of the icy pebbles, to clarify the disk conditions for SI and the runaway pile-up, which would lead to gravitational instability, of the icy pebbles (Ice-RPU) and that of the silicate grains (Sil-RPU) near the snow line in a consistent manner. We also incorporated additional potentially important physical effects that were not previously considered $-$ (1) we included the back-reactions of pebbles and silicate grains in their radial diffusion (Diff-BKR; Eqs. (\ref{eq_D_pebble}) and (\ref{eq_D_silicate})) in addition to Drift-BKR, while we neglected modulation of disk gas structure due to the solid pile-up, (2) we assumed that the released silicate grains initially have the same scale height and it increases with time by vertical turbulent diffusion up to that of the gas (Eq. (\ref{eq_silicate_H})), while \cite{Sch17} assumed it is always equal to the gas scale height and \cite{Ida16} assumed it is always equal to the pebble scale height, and (3) we distinguish between the turbulent alpha parameter ($\alpha_{\rm tur}$) and the effective alpha parameter for disk gas accretion ($\alpha_{\rm acc}$)  that could be determined by disk wind.\\

We found that Diff-BKR plays the most critical role for Sil-RPU and Ice-RPU to occur. If Diff-BKR is neglected ($K = 0$), the runaway pile-up does not occur as long as $F_{\rm p/g} < 0.6$, although the equilibrium state satisfies the condition for occurrence of SI of icy pebbles for some range of disk parameters.  For example, the solid-to-gas ratio of icy pebbles outside the snow line becomes larger than unity for the pebble mass flux $F_{\rm p/g} >  0.5$ when $\alpha_{\rm tur} < 3 \times 10^{-3}$ and $\alpha_{\rm acc}= 10^{-2}$ (Figure \ref{fig_K1K2}). The solid-to-gas ratio of silicate grains inside the snow line becomes larger than unity (a favorable condition of GI) even for small pebble mass flux $F_{\rm p/g} > 0.3$ when $\alpha_{\rm tur} < 3 \times 10^{-3}$ and $\alpha_{\rm acc}=1\times10^{-2}$ (case of $K=0$ in Figure \ref{fig_K1K2}). This threshold value is much lower than the previous work \citep{Sch17} because of the neglection of Drift-BKR of silicate grains. If Diff-BKR is included ($K = 1$, $2$), the radial diffusion to smooth out the solid concentration is weakened as the concentration proceeds. In this case, we found that either of Sil-RPU or Ice-RPU occurs, depending on the value of $\alpha_{\rm tur}/\alpha_{\rm acc}$ (Figure \ref{fig_K1K2}); Sil-RPU (that would result in formation of rocky planetesimals) is favored for $\alpha_{\rm tur}/\alpha_{\rm acc} \ll 1$ and Ice-RPU (formation of icy planetesimals) is favored for $\alpha_{\rm tur}/\alpha_{\rm acc} \sim 1$. For $\alpha_{\rm acc}= 10^{-2}$ and $K=2$, the runaway pile-up occurs for $F_{\rm p/g} > 0.2$. The threshold value of $F_{\rm p/g}$ is lower than that for satisfying the condition of the streaming instability.\\

In summary, the back-reactions of solids in radial diffusion by turbulence play critical roles on the runaway pile-up of silicate grains and icy pebbles around the snow line. Both inside and outside the snow line, solids can pile up significantly under the reasonable conditions discussed above, which would lead to formation of rocky (dry) and icy (wet) either planetesimals inside and outside of the snow line, respectively, depending on the disk evolution stage. As discussed in section \ref{sec_planet_formation}, the rocky planetesimal formation is likely when the snow line migrates to inner disk regions, while the icy planetesimal formation is likely when the snow line is still located in relatively outer disk regions. This could provide favorable initial radial distributions of rocky and icy planetesimals for our Solar system. As listed in section \ref{sec_caveats}, our present simulation still has some caveats. More detailed simulation is required to confirm our conclusions.


\begin{acknowledgements}
R.H. acknowledges the financial support of JSPS Grants-in-Aid (JP17J01269, 18K13600). S.I. acknowledges the financial support (JSPS Kakenhi 15H02065), MEXT Kakenhi 18H05438).
\end{acknowledgements}

\bibliography{planetesimals}

\end{document}